# Status of the Top Plate and Anticryostat for High Field Cable Test Facility at Fermilab

V. Nikolic, G. Velev, R. Bruce, T. Tope, D. Orris, X. Yuan and M. Kifarkis

*Abstract*— **Fermi National Accelerator Laboratory (Fermilab) is currently constructing a new High Field Vertical Magnet Test Facility (HFVMTF) designed for testing High Temperature Superconducting (HTS) cables under high magnetic fields. This facility is expected to offer capabilities similar to those of EDIPO at PSI and FRESCA2 at CERN. The background magnetic field of 15 T will be generated by a magnet supplied by Lawrence Berkeley National Laboratory. The primary function of HFVMTF will be to serve as a superconducting cable test facility, facilitating tests under high magnetic fields and a broad spectrum of cryogenic temperatures. Additionally, the facility will be utilized for testing high-field superconducting magnet models and demonstrators, including hybrid magnets, developed by the US Magnet Development Program (MDP). This paper provides a comprehensive description of the current status of two pivotal components of the facility: the Top/Lambda Plates Assembly and the Anticryostat for the Test Sample Holder. The latter will serve as a principal interface component connecting cable test samples with the facility's cryostat.**

## I. INTRODUCTION

THE US Department of Energy Offices of Science, specifically the High Energy Physics (HEP) and Fusion Energy Sciences (FES) programs, have collaboratively initiated the construction of a facility dedicated to testing High Temperature Superconductor (HTS) cables. This facility, named HFVMTF, is currently in the construction phase at Fermilab. Engineered to match or exceed the capabilities of European test stands such as EDIPO at PSI and FRESCA2 at CERN in Switzerland, HFVMTF will serve as a crucial resource for the U.S. FES and HEP communities. Its primary objective is to act as a testing platform for HTS cable samples, subjecting them to high dipole fields and elevated temperatures [1]- [3].

Within the framework of the U.S. HEP Magnet Development Program (MDP) [4], HFVMTF will play a central role as the primary testing facility for magnets generating fields in the range of 20T and beyond. This includes hybrid magnets that integrate both low-temperature and high-temperature superconductors. These advanced magnet configurations mark a significant step toward achieving dipoles with field strengths exceeding 20T, a critical element for upcoming hadron-hadron colliders.

This paper provides a detailed overview of the status of the principal interface of the magnet and the Test Sample Holder with the facility cryostat. This encompasses details related to the designs of the Top Plate Assembly, Lambda Plate Assembly, and Anticryostat for the Test Sample Holder.

## II. TOP PLATE ASSEMBLY AND ANTICRYOSTAT PARAMETERS

Based on the interface with cryostat and magnet as well as requirements from the community of users, the main parameters of Top Plate assembly and Anticryostat were selected. Some of the parameters are driven from basic requirements of this test facility described in documented [5], [6].

Since the main requirement of cryostat is to cool the magnet up to superfluid helium at a temperature of 1.9K, this requirement imposes the use of double-bath vessel with the Lambda Plate that will separate 4.5K normal liquid helium and pressurized superfluid helium at 1.9K and 1.2bar. The maximum stored energy in the magnet is on the order of 20MJ and in quench sceneries, some of the energy is to be released through Lambda Plate. The Top Plate assembly is designed to mainly support FES experiments, however secondary Top Plate will be built to support MDP experiments. Table 1 summarizes main parameters of Top Plate Assembly and Anticryostat.

In the Anticryostat a sample holder with the HTS test sample will be inserted into the magnet aperture. Its conceptual design is similar to the cryostat at EDIPO [1], [5]. The design goal is to control the sample temperature within an approximately 1K step in the region of 4.5K to 50K. To excite HTS test samples with a current up to 100kA, an SC transformer will be used, a solution that has been implemented in the SULTAN [9] and EDIPO [1] test facilities.

Submitted for manuscript September 19, 2023.

This work was supported by the US Department of Energy Offices of Science, High Energy Physics, and Fusion Energy Sciences, and the Fermi Research Alliance, LLC, under contract No. DE-AC02-07CH11359 with the US Department of Energy (Corresponding author: V. Nikolic.)

V. Nikolic, G. Velev, R. Bruce, T. Tope, D. Orris, X. Yuan and M. Kifarkis are with the Fermi National Accelerator Laboratory, Batavia, IL 60510, USA (e-mail: vnikolic@fnal.gov). Color versions of one or more of the figures in this paper are available online at http://ieeexplore.ieee.org.

Color versions of one or more of the figures in this article are available online at http://ieeexplore.ieee.org

## III. Top Plate Assembly Design

The Top Plate Assembly for HFVMTF is in final design process and will go through procurement search for manufacturing contractor in Fall of 2023. Expectation is to have this component for initial zero magnet testing and commissioning of this test stand in Fall of 2024. As for the Vertical Magnet Test Facility (VMTF) at Fermilab, the magnets tested in the new HVMTF cryostat will be supported/suspended from Top Plate Assembly using a custom designed 3-point rod support system. The Top Plate Assembly is composed of 5 mayor components, the Top Plate, the Support Rods, the 80K Shield, the Lambda Plate, and the Adapter Plate (Figure 1).

TABLE I
HFVMTF TOP PLATE ASSEMBLY PARAMETERS

| Parameter | Value |
|---|---|
| Maximum Weight of Magnet for Background Field | 20 US ton |
| Minimum Length of Magnet for Background Field | 3.200m |
| Maximum Diameter of Magnet for Background Field | 1.375m |
| Maximum Current of Magnet with HTS Leads | 20kA |
| Minimal Temperature of Magnet for Background Field | 1.9K |
| Maximum Anticryostat Diameter (above $\lambda$-plate) | 500 mm |
| Maximum Anticryostat Diameter (below $\lambda$-plate) | 140 mm |
| Maximum Cable Test Sample in Anticryostat | 3.3m |
| Cable Test Sample Temperature | 4.5-50.0K |
| Maximum Cable Test Sample Current (direct) | 16kA |
| Maximum Cable Test Sample Current (transformer) | 100kA |

The Top Plate is designed per ASME BPVC code and will be code stamped to satisfy the Fermilab and US requirements. The Top Plate could host up to four 30kA vapor cooled leads or four pairs of HTS 10kA leads. Flange adapters will be used to switch from one lead style to other with maximum lead diameter of 8.25" (210mm). The ability to install four leads in Top Plate Assembly might be required in MDP R&D tests and therefore the design of this plate shall satisfy this requirement and keep enough space for the instrumentation. However, only 2 pairs of 10kA HTS leads will be used during the tests of the FES cables and the other leads will be removed and their flanges will be blanked off. The reason for using the HTS leads is to minimize the helium consumption and minimize the operational cost. Respectively, there will be ability to use up to 4 CLIQ leads. Additionally, there will be two sets of instrumentation tree used for data acquisition, quench protection and quench characterization. This set of instrumentation tree will be able to pull up to 1,200 channels from magnets using Mill Spec connectors and new fiber optic technology for strain mapping.

The Support System that will carry the magnet will consist of three 2" stainless steel rods going from Top Plate to Adapter Plate. This support system will compensate the thermal contraction of the materials and therefore the dead load of magnet will be transferred from Top Plate to Lambda Plate during the installation process in cryostat. In this test facility, the magnetic shielding has already been installed in the pit [6] and in case of axial misalignment of magnet, the additional lateral forces may be applied on the entire Top Plate structure. Consequently, the support rods below Lambda Plate will have a calibrated strain gauges installed on them to monitor the magnet movement during powering.

The Top Plate Assembly also serves as a critical component to minimize the heat load and improve the thermal performance of the cryostat. This assembly is composed of a 80K Thermal Shield that consist of copper flat plate with Liquid Nitrogen lines mechanically attached to this plate. Above and below this plate, copper clad G-10 baffles were used with closed cell Rohacell foam as volume displacer and additional thermal barrier (Figure 1).

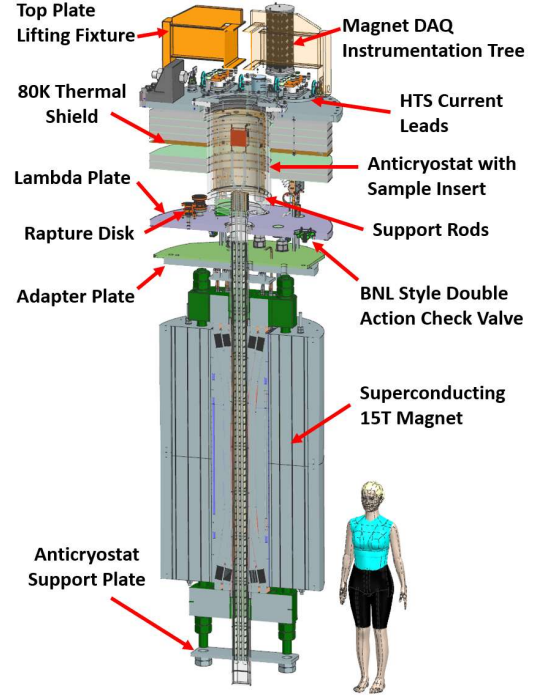

Fig. 1. Cross-section of the Top Plate Assembly model for Test Facility with magnet and Anticryostat

Unlike other cold test stands at Fermilab that use Lambda Plate made out of G-10 for separation of 1.9K and 4.4K region, the Lambda Plate in HFVMTF will be stainless steel to satisfy the ASME BPVC code requirements, although it will not be ASME code stamped. To minimize the heat load on this large diameter plate, a thin G-10 plates will be using Stycast© on top and bottom side of stainless steel Lambda Plate. Unlike most double-bath test stands, a surface-to-surface contact seal which consist of a spring-energized seal [7], [8] is preferred to minimize the leaks at the location of the lambda ring. This solution has already been tested at Brookhaven National Laboratory. From the safety point of view, to accommodate the pressure across the Lambda Plate during the cooling process and the quench of the magnet there will be a check valve designed by BNL. This valve will be installed on the bottom side of the Lambda Plate to minimize heat load. The BNL valve is indeed composed of two check valves. The primary check valve will serve to protects the superfluid bath from over pressurizing during the quench of the magnet. The secondary valve, called the reverse check valve, will accommodate the pressure between the two baths during the cool down from 4.5K to 1.9K as the helium density increases between these two temperatures and the pressure will drop in this volume. As a protection from catastrophic failures, two rupture disks will also be installed on the Lambda Plate. They will both be reverse buckling disks with a burst pressure of 15psi (1bar) at 4.5K and will resist full vacuum. This rupture disks are not supposed to burst during quench, but high-pressure peaks can appear in the lower vessel during this sudden event and partially damage the equipment and therefore this check valves will serve to protect from worst-case scenario, for example, loss of insulating vacuum in the cryostat, check valve fails to open, etc. Although these two devices will add heat loads to the superfluid bath, they will serve as a secondary protection system and satisfy code requirements. Additionally, there will be plugs made from G-10 for instrumentation and bus feedthrough that will be potted using Stycast©. Center opening will be used for Anticryostat feedthrough and appropriate Teflon base energized c piston type seal will be used with flange to connect to Lambda Plate.

The Adapter Plate is introduced between Lambda Plate and Magnet to allow as easy interface on this Top Plate Assembly with various Magnets that will be installed in HFVMTF (Figure 1). Adapter plate will be made from 2" thick stainless steel with thin layer on top from of G-10 to serve for dielectric insulation of spliced leads that are typically mechanically connected to adapter plate to sustain Lawrence (magnetic) forces.

Lastly, the Anticryostat placed in aperture of the magnet, will be attached to Top Plate Assembly, and have opening access on top for sample insertion.

## IV. TOP PLATE INSTRUMENTATION

The two superconducting magnet systems, each with its own branch of power supplies and quench protection systems, also have their own branch of instrumentation for monitoring and controls. These include quench detection and quench characterization voltage tap wiring, cryogenic temperature sensors, magnet strain gauges, cold pressure transducers, acoustical transducers, quench antenna, and liquid level sensors. There will also be feedthroughs for HTS fiber optic quench detection sensors. In addition, each magnet system will have current source and voltage sense wiring for the magnet protection heaters, and there will be superconducting bus and power leads for the coupling-loss induced quench (CLIQ) systems [10]. The magnet and bus instrumentation sensors along with all other instrumentation will interface via their respective instrumentation trees with multipurpose-signal feedthrough capability. Each superconducting magnet test system branch will support the following instrumentation: 32 channels each of Primary and redundant quench signals with independent VTs (Whole Coil, Half-coils, Magnet SC bus, Test stand SC Bus, Vapor-cooled Cu power leads with flag connection monitoring); Three chassis of quench characterization modules for a total of 96 channels; Feedthrough wiring for Liquid level sensors to sense liquid helium level below and above the Lambda Plate; Feedthrough wiring for cold pressure transducers to sense differential pressure across the Lambda Plate; Feedthrough signals for up to 24 magnet strain gauges; Feedthrough signals for up to 10 acoustical sensors; Feedthrough wiring for up to 100 quench antenna channels; Fiber-optic quench signals – total feedthroughs depend on technology; Current wiring and heater voltage tap wires for 4 Protection heater supplies; Current wiring for 2 spot heater supplies; Superconducting bus and copper leads for 1 CLIQ system.

## V. ANTICRYOSTAT AND TEST SAMPLE HOLDER

The Anticryostat and Test Sample Holder are in preliminary design stage, expecting to start finalization of this design after procurement of Top Plate Assembly design, which based on milestones it to be delivered at Fermilab by end of summer 2024.

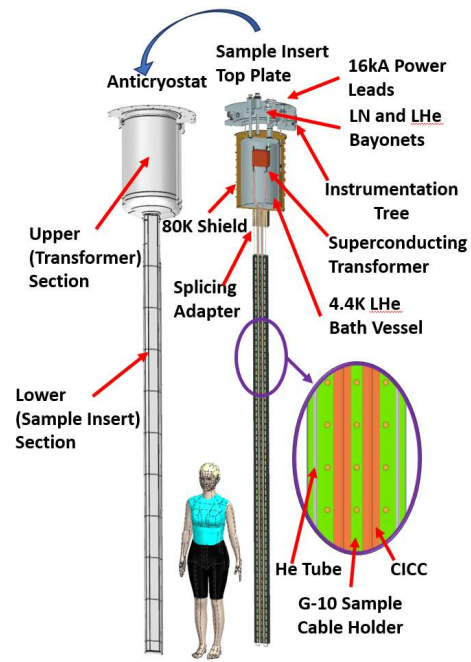

Fig. 2. Cross-section of the preliminary Anticryostat cryostat design and Sample Insert with Transformer Assembly.

The Anticryostat is designed to hold a sample, a superconducting transformer, and process piping (Figure 2). To avoid unnecessary warm up and cooling of the magnet providing the background field, which will significantly reduce the operational cost, Anticryostat is designed so that the sample holder can be removed and reinserted while maintaining the magnet at cold environment. The Anticryostat is designed to operate as a vacuum vessel specifically designed to hold environment for controlled temperatures of sample from 4.5K to 80K while operation per user specification. To achieve this, the Anticryostat will be made from 316L stainless steel inner and outer rectangle profiled tubes, where in interspace between the tubes will be MLI in vacuum space supported by spiders for minimizing thermal load from magnet space to sample space (Figure 2,3).

The sample insert is like the EDIPO sample holder design [1]. The insert will have two mayor components. In upper section will be a superconducting transformer in liquid helium bath at 4.4K, which will require to have a small helium vessel and an 80K shield surrounding the vessel to minimize the thermal load. Below the transformer helium vessel, there will be an adapter that will connect the superconducting transformer leads to HTS cable sample.

The second mayor component is sample cable holder (Figure 2,3). The profile of sample cable holder that will be inserted in Anticryostat/magnet aperture is rectangular, as the one in magnet, and will be suspended along with transformer from the Anticryostat Top Plate Assembly. Sample Cable Holder will be made from G-10 for magnetic, conductive, and mechanical properties and will be used to hold cable at desired spacing and position inside the magnetic field while withstanding all forces generated under ultimate magnetic flux which are estimated to be 340kips (1.5MN) at background field of 15T.

From collaboration meetings, the new user requirements introduced a different configuration of length for Sample Cable Holder to accommodate the joint offset in field and therefore the appropriate spacers will be added to accommodate this offset. Along the sides of the Sample Cable Holder will be a capil-

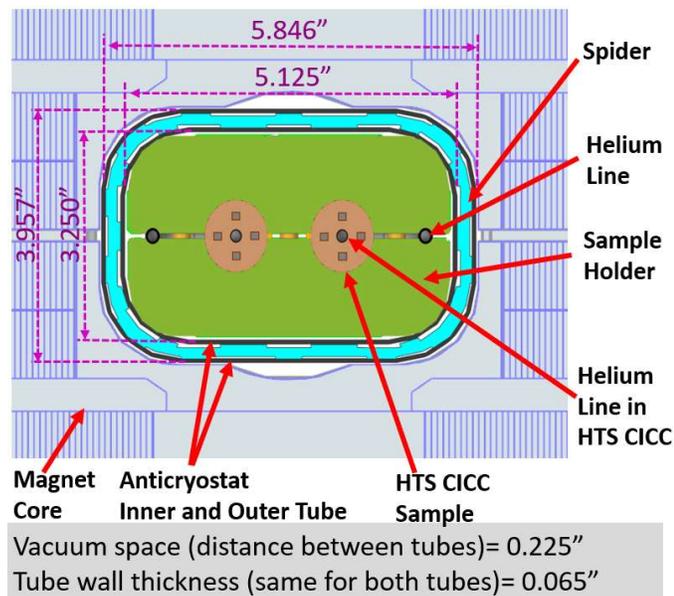

**Spider**

**Helium Line**

**Sample Holder**

**Helium Line in HTS CICC**

**Magnet Core**   **Anticryostat Inner and Outer Tube**   **HTS CICC Sample**

Vacuum space (distance between tubes)= 0.225"
Tube wall thickness (same for both tubes)= 0.065"

Fig. 3. Cross-section of the preliminary design of the Anticryostat cryostat with the sample inside and cable in conduit conductor (CICC).

lary 3/16" (4.7mm) ID tubes that will supply a forced flow 4.5K-50K gas helium at programmable mass flow rate up to 145psi (10bar). Additionally, these capillary tubes will have the ability to be used as supply or return in parallel arrangement or combination of supply and return in series connection of cable sample arrangement (Figure 3).

During the R&D test phase, the Anticryostat, also known as the Warm Bore, will utilize magnets similar to those currently in use at other Magnet Test Facilities within Fermilab. The Warm Bore is primarily employed for capturing warm magnetic measurements or conducting quench studies utilizing various probes. It typically accommodates two sizes of warm fingers: 5" (127mm) OD and 2.5" (63mm) OD. Its internal space is anticipated to function under vacuum or room temperature conditions while performing magnetic measurements. However, the intricate design specifics for this operation are slated to be finalized by 2025.

## VI. Conclusion

In US, the Fermilab is in process of building a new cable-testing facility that will have similar capability of the European cable test facilities EDIPO and FRESCA2. This test facility primarily will serve the communities within the DOE Office of Science, the US Magnet Development Program, and the US Fusion Energy Sciences programs, to test HTS cable samples in a magnetic field up to 15T.

This paper reports on the progress of the design and construction of main interface components in facility. The cryostat is in the fabrication process. The Top Plate Assemblies has final engineering design and will soon begin the search of manufacturing vendor for fabrication later this year. From community of user's workshop, the Anticryostat and sample cable holder preliminary design is complete and will soon undergo a review before it gets finalized.

## References

[1] P. Bruzzone *et al.*, "EDIPO: The Test Facility for High-Current High-Field HTS Superconductors," *IEEE Trans. Appl. Supercond.*, vol. 26, Art. no. 9500106 (2016).

[2] A.P. Verweij *et al.*, "1.9 K test facility for the reception of the superconducting cables for the LHC," *IEEE Trans. Appl. Supercond.*, vol. 9, Issue 2 (2021).

[3] V. Benda *et al.*, "Cryogenic design of the new high field magnet test facility at CERN," *Physics Procedia 67 (2015) 302 – 307*

[4] S. Prestemon *et al*., "The 2020 Updated Roadmaps for the U.S. Magnet Development Program," https://science.osti.gov/hep/Community-Resources/Reports

[5] G. Velev *et al*., "Design and Construction of a High Field Cable Test Facility at Fermilab," *IEEE Trans. Appl. Supercond.*, vol. 31, Art. no. 9500304 (2021).

[6] G. Velev *et al*., "Status of the High Field Cable Test Facility at Fermilab," *IEEE Trans. Appl. Supercond.*, vol. 33, Art. no. 9500306 (2023).

[7] R. Bruce *et al*., "Cryogenic and safety design of the future High Field Cable Test Facility at Fermilab," FERMILAB-CONF-23-354-TD, *Web. doi:10.2172/1989882*

[8] S. Koshelev et al., "Design of the cryostat for high field vertical magnet testing facility at Fermilab," in Proc. IOP Conf. Ser.: Mater. Sci. Eng., 2022, Art. no. 012081.

[9] T. J. Peterson, R. J. Rabehl, and C. D. Sylvester, "A 1400 liter 1.8K test facility," Adv. Cryogenic Eng., vol. 43, pp. 541–548, 1998.

[10] J. Elen *et al.*, "The superconductor test facility Sultan," *IEEE Trans. on Magnetics*, vol. 17, issue 1. (1981).

[11] A. Galt *et al.*, "A Quench Detection and Monitoring System for Superconducting Magnets at Fermilab," *IEEE Trans. Appl. Supercond.*, vol. 32, Art. no. 9500404 (2022).